\documentclass[letterpaper, 10pt, journal, twoside, pdftex]{IEEEtran}      
\usepackage{graphicx}
\usepackage{epsfig}
\usepackage{bm}
\usepackage{amsmath}
\usepackage{amsfonts}
\usepackage{amssymb}
\usepackage{ascmac}
\usepackage{prettyref}
\usepackage{comment}
\usepackage{multirow}
\usepackage{makeidx}
\usepackage{fancyhdr}
\usepackage{color}
\usepackage[caption=false,font=footnotesize]{subfig}
\usepackage{float}
\usepackage{cite}
\usepackage{algpseudocode,algorithm}

\usepackage{color}

\newrefformat{sec}{\ref{#1}“\nameref{#1}”}
\newrefformat{sub}{\ref{#1}“\nameref{#1}”}
\newrefformat{par}{\ref{#1}“\nameref{#1}”}
\newrefformat{tab}{\ref{#1}“\nameref{#1}”}
\newrefformat{fig}{Fig. \ref{#1}}

\IEEEoverridecommandlockouts                              

\begin{document}

\title{
Force control of grinding process\\
based on frequency analysis
}


\author{Yuya Nogi$^{1}$, Sho Sakaino$^{2}$, Toshiaki Tsuji$^{1}$,
\thanks{*This work was partially supported by JSPS KAKENHI, Japan Grant Number 21H01280. }
\thanks{$^{1}$Yuya Nogi and Toshiaki Tsuji are with the Dept. Electrical and Electronic Systems, Saitama University, 255 Shimo-okubo, Saitama, 338-8570}
\thanks{$^{2}$Sho Sakaino is with the Graduate School of Systems and
Information Engineering, University of Tsukuba, 1-1-1 Tennodai,
Tsukuba, Ibaraki 305-8577}
} 
\maketitle


\begin{abstract}
Hysteresis-induced drift is a major issue in the detection of force induced 
during grinding and cutting operations.
In this paper, we propose an external force estimation method based 
on the Mel spectrogram of the force obtained from a force sensor. 
We focus on the frequent strong correlation between the vibration frequency 
and the external force in operations with periodic vibrations. 
The frequency information is found to be more effective for an accurate force estimation 
than the amplitude in cases with large noise caused by vibration. 
We experimentally demonstrate that the force estimation method 
that combines the Mel spectrogram with a neural network is robust against drift.
\end{abstract}


\section{INTRODUCTION}
\IEEEPARstart{M}{any} 
manufacturing processes, such as assembly and grinding, 
involve contact with objects. 
The variability in the state of contact with an object 
often increases with increasing task complexity.
The automation of tasks with changing contact states, 
which are called contact-rich tasks, 
has been widely studied in robotics~\cite{Walker1,Kakinuma2,Polden3}.
In contact-rich tasks, the contact force varies considerably with the contact state changes. 
In such tasks, small fluctuations in the force must be detected to determine whether contact is possible. 
In addition to the detection of large forces, 
the extraction of features from low forces requires force measurement 
with a wide dynamic range~\cite{Xu4}; hence, force sensing is an important 
technology influencing the performance of contact-rich tasks. 
Force sensing often limits the performance in simple tasks;
wherein few changes occur in the contact state. 
For example, in grinding, the magnitude of the contact force significantly 
affects the quality of the product surface~\cite{Roswell5}, 
and it is desirable to control the contact force to a constant or desired 
magnitude. Force control techniques, such as hybrid control~\cite{Raibert6} 
and impedance control~\cite{Hogan7}, are widely used to adjust the force 
to a desired value in tasks involving contact. 
Attempts have been made to improve the performance of these 
methods~\cite{Kiguchi8,Schindlbeck9}. 
The most important factor determining the 
performance of such control methods is force sensing. However, problems, such as 
hysteresis, temperature drift, and deviation of the offset values in the working 
posture, cause steady-state errors in the force information. In addition, 
there is fluctuation in the obtained force data because of the vibration of the tool 
attached to the end-effector (EEF) during grinding and cutting. 
If such force information is used in the control system without 
considering these problems, chattering may occur, and force control 
may become unstable.

Hence, attempts to improve force control using force information
with less noise are underway. The external force estimation observer 
is a technique for estimating the force from the joint torque using 
an observer instead of acquiring it using a force sensor~\cite{Murakami10}; 
however, noise caused by static friction is an issue when a geared 
motor is used. The force information can be  estimated from the 
joint torque applied to a grinding work~\cite{Yunfei11}, and many 
attempts have been made to estimate the joint torque from series 
elastic actuator displacement and to use it for force control \cite{pratt12,kong13,kong14}. 
Unfortunately, in methods that infer the force from the joint torque 
of the robot, the inertial forces of the links connected in series cause noise. 
Despite the attempts made to improve the accuracy of force sensors~\cite{Okumura15}, 
the hysteresis of six-axis force sensors may occur up to approximately 1\% 
of the full scale, which limits the detection of fine contact.

In this study, we focus on the frequent strong correlation between the 
vibration frequency and the external force in operations with 
periodic vibrations, such as grinding and cutting. 
The frequency information is found to be more effective than the amplitude 
for accurate force estimation in noisy tasks. 
In addition, it is more robust against problems specific to force 
sensors, such as hysteresis. Although the correlation between 
the force information and the frequency components is strong, 
the relationship is highly nonlinear. Therefore, we introduce a 
neural network (NN)-based machine learning method for inference. 
Many attempts have been made to improve force estimation 
and force control using NNs~\cite{ananthan16,piao17}; 
however, none have shown that the control performance can be 
improved by combining them with a real-time frequency analysis.
Through practical experiments, we evaluate the accuracy of force 
estimation and the performance of the grinding operation on 
a flat surface using a six-degree-of-freedom (6-DOF) manipulator 
and demonstrate the effectiveness of the proposed method. 

The remainder of this paper is organized as follows. Section \ref{METHODS} 
describes the principle of the proposed method. Section \ref{EXPERIMENT} 
presents the experiments and their results, demonstrating the effectiveness 
of the proposed method. Finally, Section \ref{conclusion} summarizes this paper.
 
\section{methods}\label{METHODS}
In this section, we first describe the experimental setup and the verification 
task, namely grinding, to clarify the preconditions for this study. 
Next, the characteristics of the force information and frequency analysis 
of the grinding process are described. Finally, the machine learning model is discussed.

\subsection{Prerequisites}
\begin{figure}[t]
  \centering
    \includegraphics[width=7.0cm]{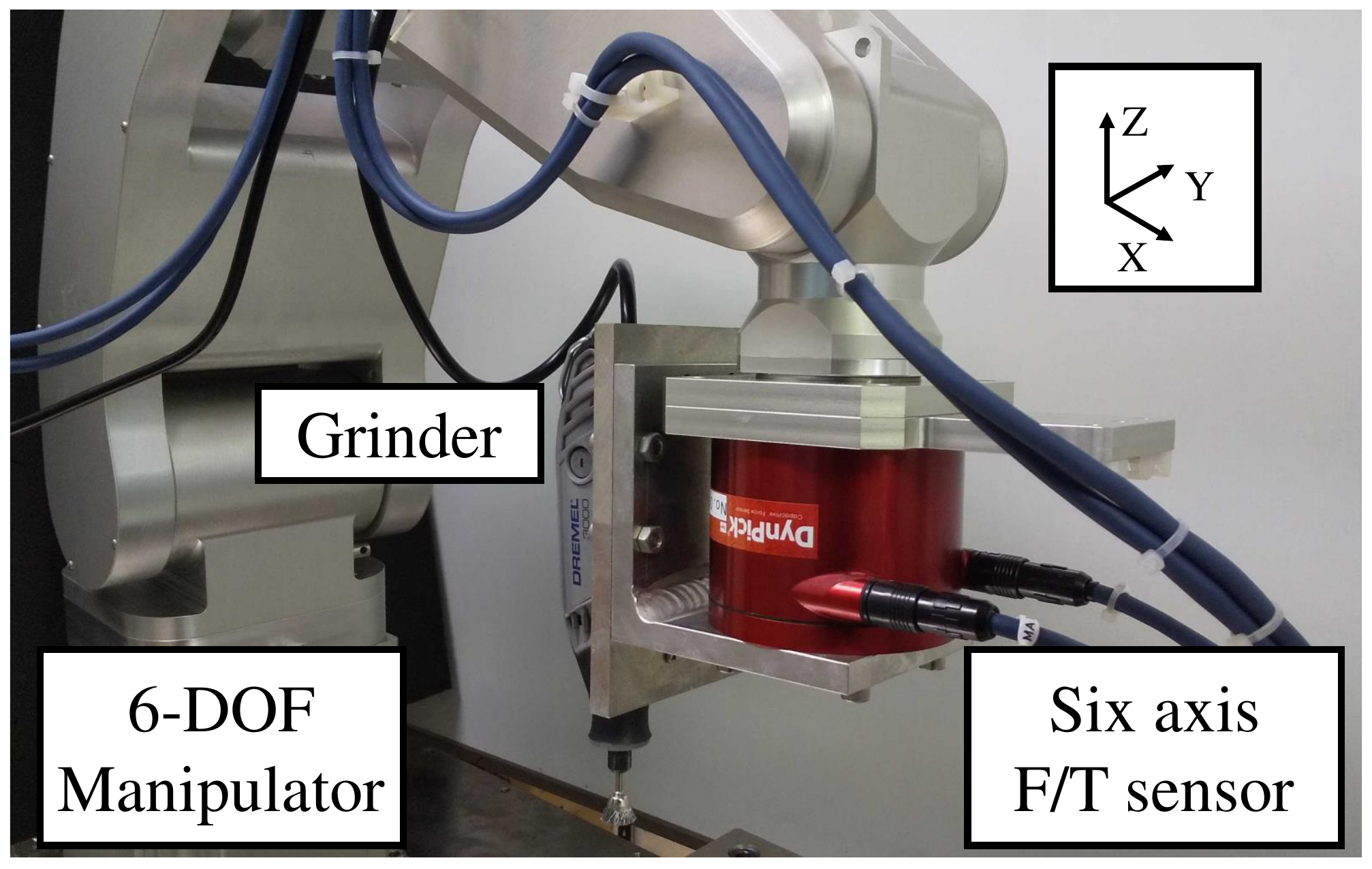}
    \caption{Experimental setup.}
    \label{fig:robot}
\end{figure}

\begin{figure}[t]
  \centering
    \includegraphics[width=7.0cm]{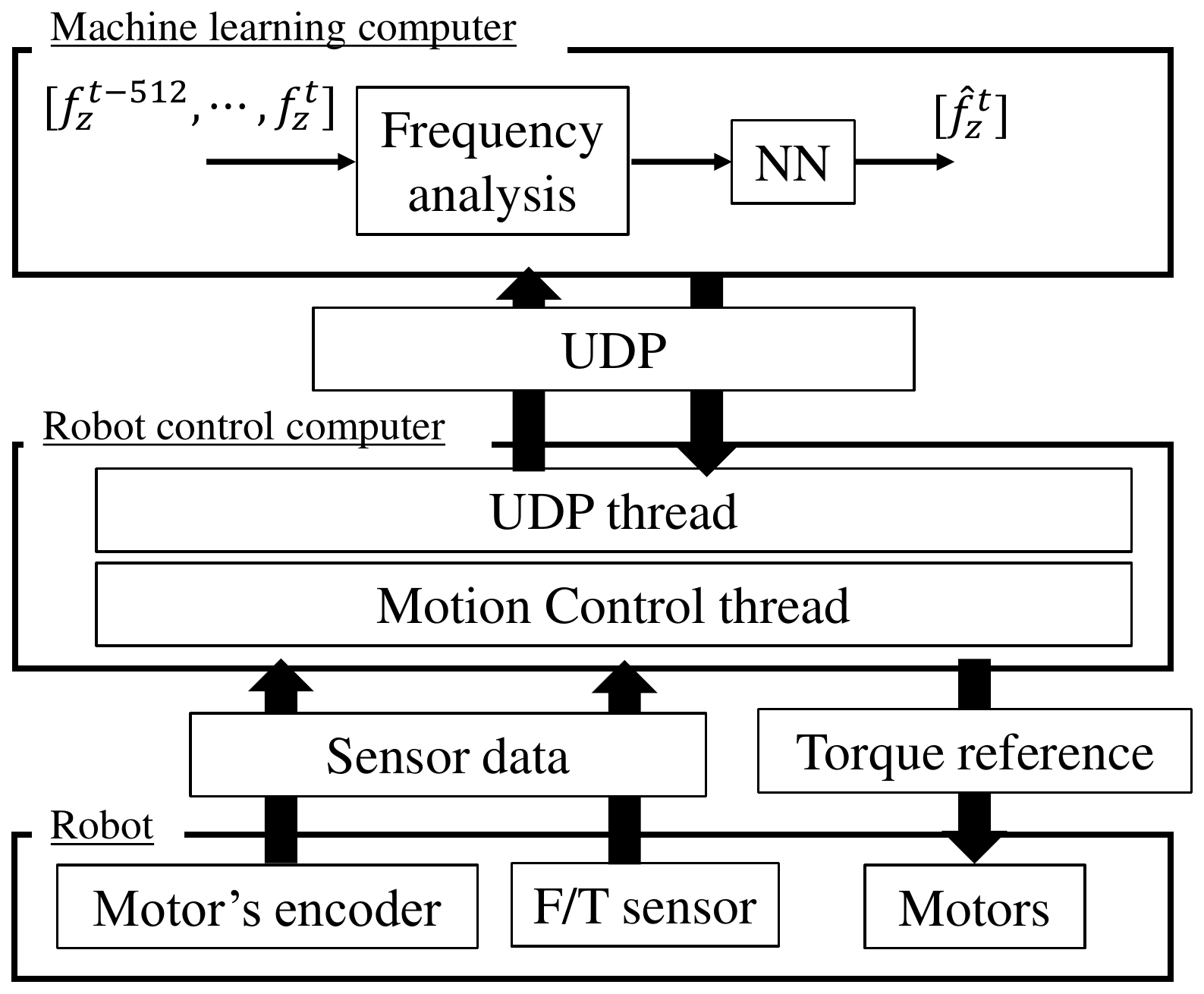}
    \caption{System architecture.}
    \label{fig:architecture}
\end{figure}

Fig.~\ref{fig:robot} shows the robot used in this study. 
A 6-DOF manipulator, equipped with a six-axis force/torque (F/T) 
sensor and a grinder as a tool for grinding, was used in the experiments. 
The grinder was not equipped with additional sensors (e.g., speed sensor); 
thus, the experiment could be performed with a general industrial robot 
setup. Fig.~\ref{fig:architecture} shows the system configuration of the 
proposed method. Two PCs for control and force prediction were used. 
User Datagram Protocol (UDP) communication was 
used between the control PC and the force prediction PC to send and receive data.
The control PC had an Intel Core i7-10700K CPU, 
and the force prediction PC had an Intel Core i9-10989XE CPU and 
a GeForce GTX1660 SUPER GPU.

\begin{figure}[t]
  \centering
    \includegraphics[width=7.0cm]{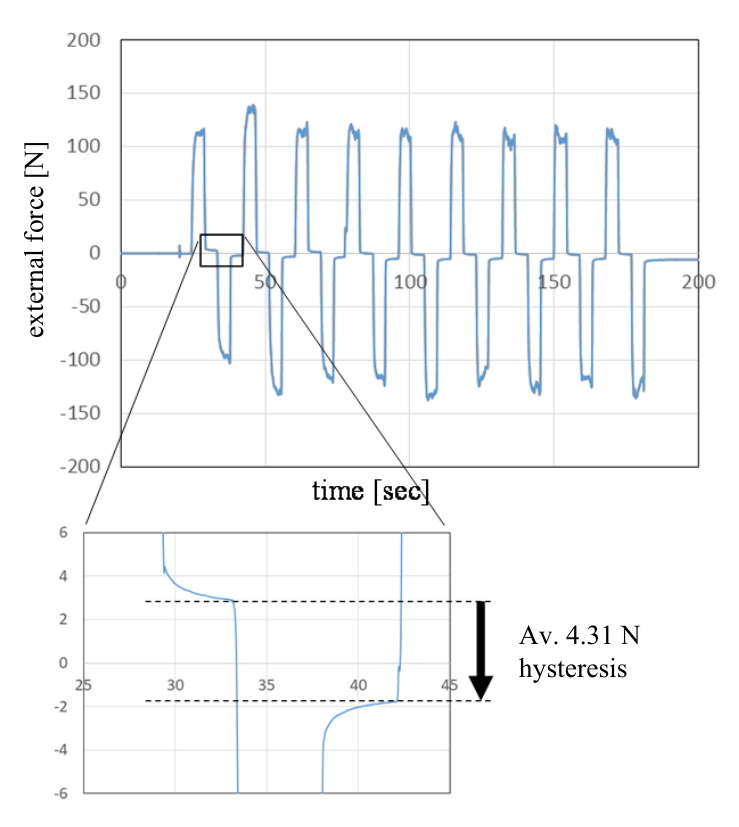}
    \caption{\textcolor{black}{Force response and hysteresis in the experimental machine used.}}
    \label{fig:hystereisis}
\end{figure}

Fig. \ref{fig:hystereisis} shows the force response of the force sensor 
when it was pushed up and down by hand nine times alternately. 
A steady response remained even during the 
period when the external force was removed after the push. 
This steady response is due to the hysteresis of the force sensors, 
which is the main cause of error in various types of force sensors. 
In this test, the average value of the hysteresis after applying a force 
of approximately 100~N was 4.31~N, and its variance was 1.54~N.
\textcolor{black}{
These results show that grinding with this robot may result in 
an error of a few Newtons due to hysteresis.
} 

\subsection{Concept of this study}
\begin{figure}[t]
  \centering
    \includegraphics[width=6.5cm]{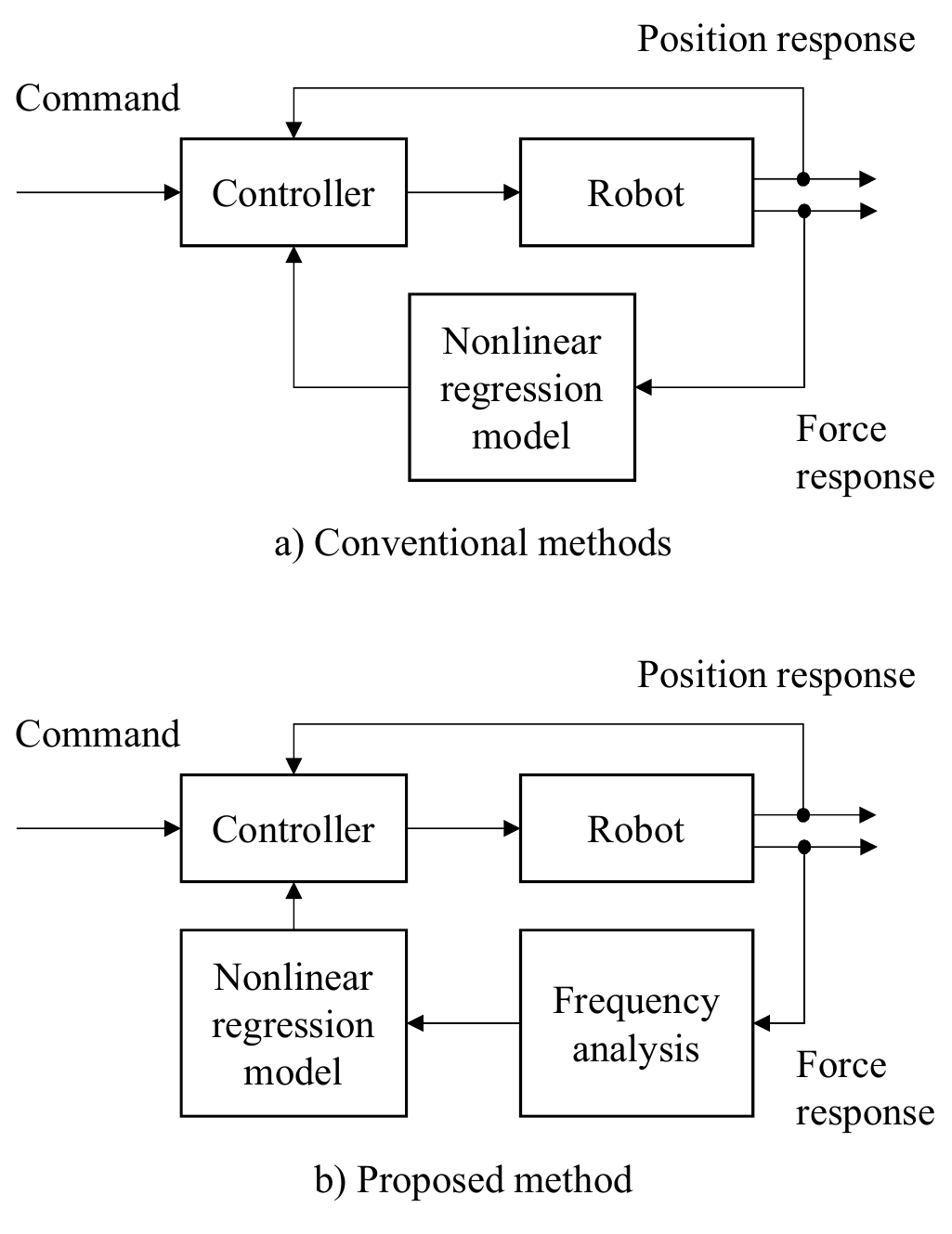}
    \caption{Concept of this study.}
    \label{fig:concept}
\end{figure}

Grinding is adopted as a typical task in which the force fluctuates because of the influence of the tool used. 
Fig.~\ref{fig:concept} shows the concept of this study. In conventional 
force control, the information from the force sensor is used as it is or is processed using 
a nonlinear regression method (e.g., low-pass filter (LPF)) to improve the force estimation accuracy
and the performance of the feedback control. In this study, we show that the varying 
force information can be converted into more accurate force information 
by adding a frequency analysis to the nonlinear regression process.
A NN-based machine-learning method was used as the nonlinear regression model. 


For the evaluation, we also adopted the force information obtained from 
a worktable-type force sensor~\cite{Totsu24} as the correct answer label. 
Because this sensor is an installed type, it is unaffected by the drift 
due to the working posture; in addition, this sensor is recalibrated 
from the response when the robot is not in contact with it to realize 
an accurate measurement. Moreover, because the sensor is not directly 
attached to the tool, it is not susceptible to vibrations. 
The sampling time of both the EEF and worktable-type force sensors was 1 ms.

\subsection{Feature Representation of Mel spectrogram}
In this study, we focused on the vibration of the force information. 
In tasks, such as grinding and cutting, the vibration of the tool also 
affects the force information. Therefore, it is necessary to extract 
features from oscillatory signals. Thus, we used the Mel 
spectrogram (MS) \cite{Shen17, Meng18} as a real-time frequency analysis. 
The MS is calculated based on the short-time Fourier transform (STFT) 
and is obtained by applying a nonlinear transformation 
to the frequency-axis of the STFT, 
which calculates the frequency information with a smaller 
number of dimensions considering human auditory characteristics. 
By deriving the MS, it is expected that the force information 
can be represented in a lower dimension. 
The Mel scale (\textit{mel}) is converted from 
the frequency (\textit{f}) using the following equation:
\begin{eqnarray}
mel=2595\log(1+\frac{f}{700})
\end{eqnarray}

\begin{figure}[t]
  \centering
    \includegraphics[width=7.0cm]{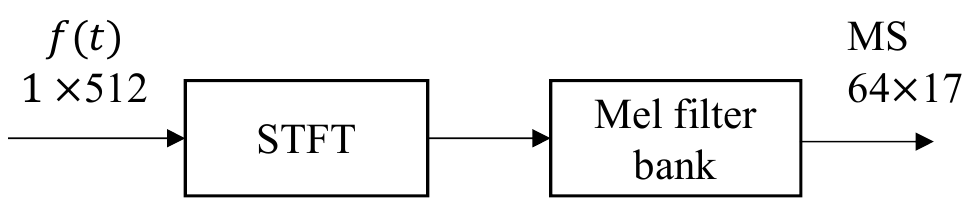}
    \caption{\textcolor{black}{Conceptual diagram of the process of calculating Mel spectrogram.}}
    \label{fig:mel-spectrogram}
\end{figure}

The number of data points that can be acquired is limited 
because of the online estimation of the force information. 
Therefore, in this study, 512 samples (512 ms) were used to calculate the MS.
Fig.~\ref{fig:mel-spectrogram} shows a conceptual diagram of the 
MS calculation, performed by STFT using a 256~ms 
frame size, 32~ms frame hop, and Hann window function. 
A 64-channel Mel filter bank was applied to the calculated STFT. 

Typically, a pre-emphasis filter is 
introduced as a pre-processing step in speech processing~\cite{Han25}. 
However, in this study, we used raw data from the force 
sensor because the bandwidth of the force information was
shorter than that of the speech.

\begin{figure}[t]
  \centering
    \includegraphics[width=8.8cm]{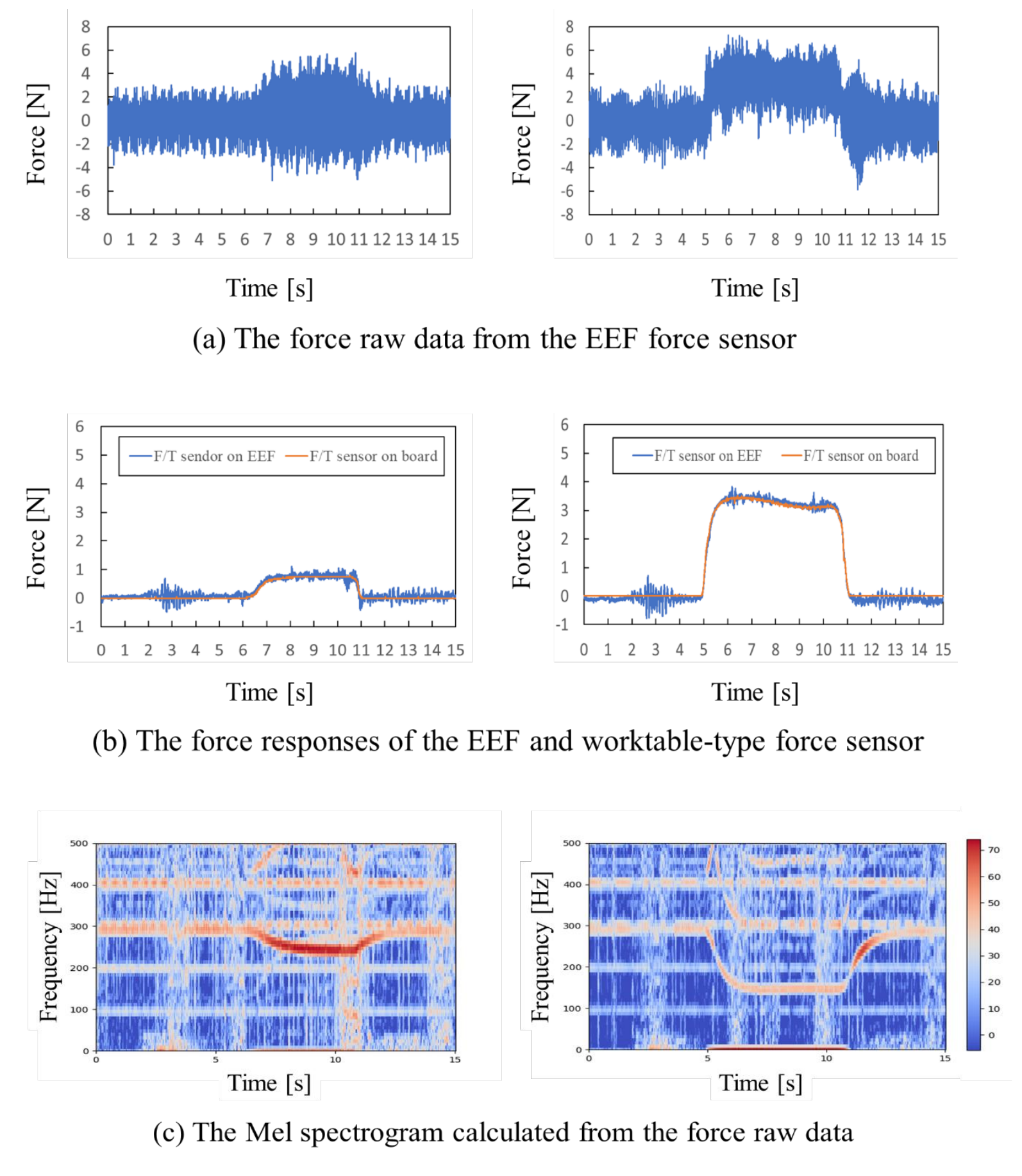}
    \caption{\textcolor{black}{
    Feature of force responses and Mel spectrogram, 
    left figures are for contact force of about 1~N, 
    right figures are for contact force of about 3~N. 
    }}
    \label{fig:feature_force}
\end{figure}

Fig.~\ref{fig:feature_force} shows the features of the force information for each force sensor.
Fig. \ref{fig:feature_force}(a) shows the force 
information of the EEF force sensor. Fig.~\ref{fig:feature_force}(b) 
shows the force information obtained by applying a first-order 
LPF with a cutoff frequency of 5 Hz to (a) 
and the force information of the worktable-type force sensor. 
Fig. \ref{fig:feature_force}(c) presents the MS
calculated from the force information in (a). 
From the results of (b) and (c), it can be confirmed that the 
MS varies with the contact force. 
Therefore, the contact force can be inferred 
using the MS as a feature input to the NN. 

\textcolor{black}{
In this method, it is important to design the range of the frequency variation.
It is detemined by the maximum frequency $f_{\rm{max}}$ and 
minimum frequency $f_{\rm{min}}$ that are allowed to change.
In general, $f_{\rm{max}}$ is determined by the sampling time [ms]
of the force sensor.
The sampling time was 1~ms; therefore, $f_{\rm{max}}$ was 500~Hz.
However, it is difficult to calculate $f_{\rm{min}}$ in advance,
because $f_{\rm{min}}$ is strongly affected by the frequency 
of the rotation speed of the tool and the hysteresis value of the force sensor. 
In addition, when the contact force exceeds a certain level, 
the rotation of the tool becomes nonuniform, 
and there is no strong relationship between the contact force and the frequency.
Therefore, it is better to set $f_{\rm{min}}$ depending on the task.
In this study, we designed the experimental machine 
such that the frequency varies between 100~Hz and 300~Hz with respect to the contact force, 
leaving a margin to avoid reaching $f_{\rm{max}}$ and $f_{\rm{min}}$.
}

\textcolor{black}{
The features to be inputted to the NN must have frequency components between 100~Hz and 300~Hz.
At the same time, since the deviations due to hysteresis 
and offset of the force sensor exist at low frequencies, 
it is necessary to remove the low-frequency components from the feature values. 
Therefore, out of the calculated 64-dimensional MS, 
we deleted the features of the bottom five dimensions and the top 14 dimensions. 
The low-frequency feature removal should be designed depending on the task.
The deletion of the 14-dimensional features above is 
to remove the effect of noise and to reduce the computational cost.
Since the 6th order corresponds to 40~Hz and the 50th order corresponds 
to 400~Hz, the features are inputted considering only the information from 40~Hz to 400~Hz.
}

\textcolor{black}{
In the field of speech recognition, STFT, MS, 
and Mel-frequency cepstral coefficient (MFCC)\cite{Martinez20, Muda19} are often used as features.
}
The difference between the STFT and MS is the presence of the Mel frequency 
transformation, and the difference between MS and MFCC is the presence of a discrete cosine transform. 
In literature \cite{Tsuji22}, determining the MFCC, which also 
includes the Mel frequency transform, is considered a promising method 
for force feature identification; thus, the performances of the three values, 
namely the STFT, the log transform of STFT (MS), and MFCC, were compared. 
The results show that the effect of the STFT conversion to the frequency 
is the most dominant, whereas the conversion to the Mel frequency 
has a slight effect. 
\textcolor{black}{
However, these results vary depending on the task.
We also conducted accuracy verification using each feature 
to verify the influence of the frequency analysis method.
}

\subsection{Prediction Network based on 1D-CNN}
Table~\ref{tb:TDNN} presents the time delay NN (TDNN) model. 
The TDNN is equivalent to a one-dimensional convolutional NN (1D-CNN) 
with chronologically ordered information and comprises two convolutional 
layers and three fully connected layers.
ReLU was used as the activation function in both the layers. 
Adam was used to update the parameters, and the initial learning 
rate was set to 0.001.

\begin{table}[t]
\centering
\caption{\textcolor{black}{TDNN based on 1D-CNN}}
\begin{tabular}{c|c|c|c}
\hline
Layer&Filter size&Stride&Output size\\ \hline \hline
Input&-&-&(17, 45)\\ \hline
Conv. 1D&3&1&	(15, 20)\\ \hline
Average Pool&2&2&(7, 20)\\ \hline
Conv. 1D&2&1&(6, 10)\\ \hline
Average Pool&2&2&(3, 10)\\ \hline
Fc&-&-&30\\  \hline
Fc&-&-&30\\ \hline
Fc&-&-&1\\ \hline
\end{tabular}
\label{tb:TDNN}
\end{table}

\section{EXPERIMENT}\label{EXPERIMENT}

\subsection{Control Architecture}
\begin{figure}[t]
  \centering
    \includegraphics[width=80mm]{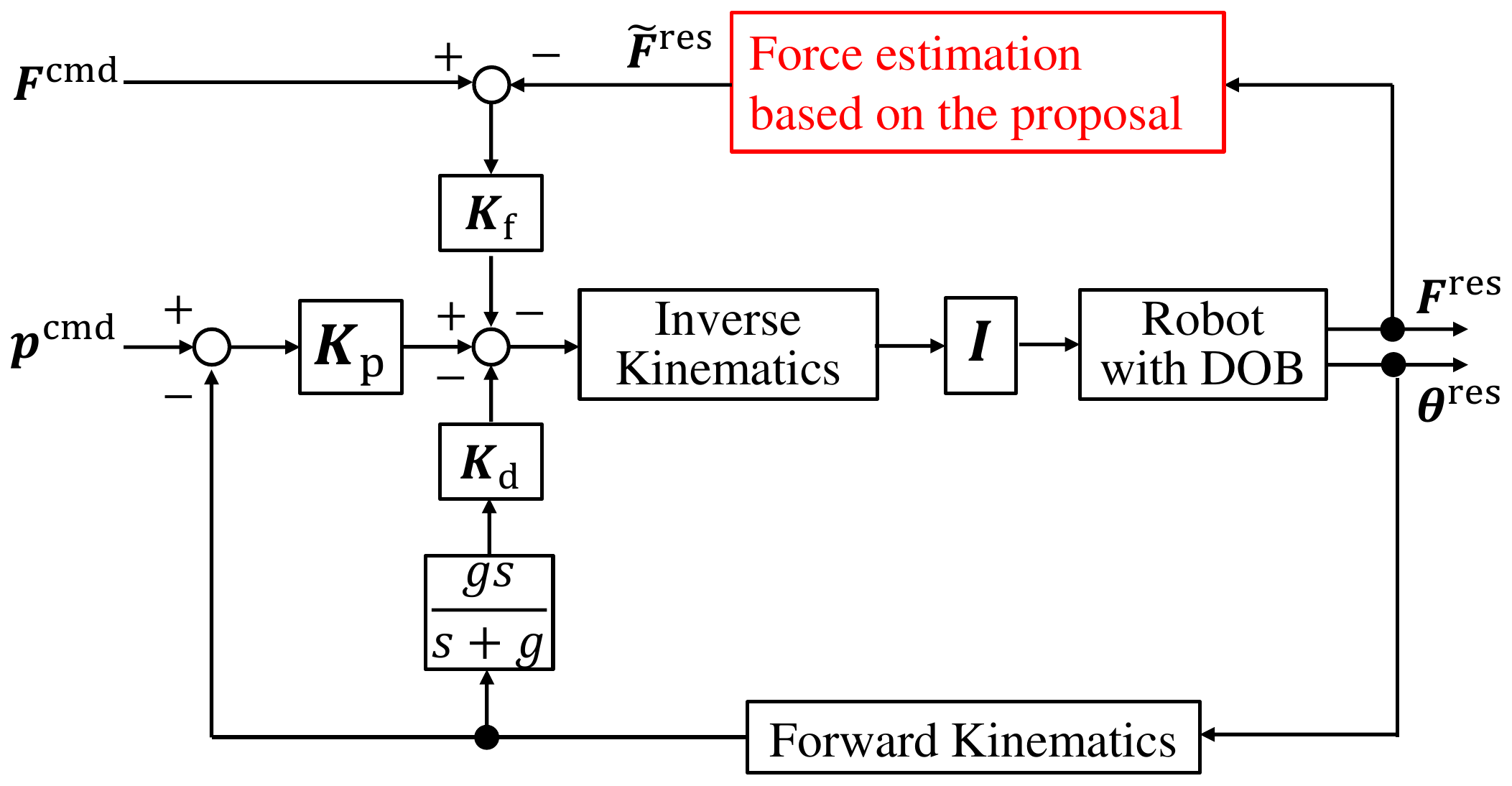}
    \caption{Block diagram of impedance control based on proposed method.}
    \label{fig:proposed_method}
\end{figure}

A general impedance control system was used; 
Fig.~\ref{fig:proposed_method} shows the block diagram of the control system. 
A disturbance observer (DOB)~\cite{Ohnishi26} was used to ensure 
the disturbance of the position and force control system. 
Table~\ref{tb:parameters} lists the control parameters. 
Here, $\bm{p}^\mathrm{cmd}$ was created by combining simple linear 
trajectories, and $\bm{F}^\mathrm{cmd}$ was taken as a constant. 
\textcolor{black}{
The proposed method is incorporated into the force control system, 
and the estimated value $\tilde{\bm{F}}^{\rm{res}}$ is outputted from the output value $\bm{F}^{\rm{res}}$ of the F/T sensor. 
}

\begin{table}[t]
\centering
\caption{Parameters of Controller}
\begin{tabular}{c|p{25mm}|p{30mm}}
\hline
$\bm{K}_\textrm{p}$&Proportional gain&
diag(700, 700, 150, 1500,\par 1500, 1500)\\ \hline
$\bm{K}_\textrm{d}$&Derivative gain&
diag(70, 70, 120, 80, 80, 80)\\ \hline
$\bm{K}_\textrm{f}$&Force gain&
diag(0.0, 0.0, 0.1, 0.0, 0.0, 0.0)\\ \hline
$\bm{I}$&Moment of inertia&
diag(1.58, 1.40, 0.80, 0.18, 0.16, 0.04)\\ \hline
$g$&Cutoff freq. of the deriv. filter&10 [Hz]\\ \hline
$T^c_s$&Sample time of the controller&
\par 0.001 [s]\\ \hline

\end{tabular}
\label{tb:parameters}
\end{table}

\subsection{Preparation of Datasets} 
\textcolor{black}{
To verify the generalization performance of the model, we obtained datasets for four cases.
Table~\ref{tb:Dataset} presents the data acquisition conditions for each dataset.
Data1 is a dataset of a simple grinding task without movement of the X-Y plane. 
In this dataset, as shown in Fig. \ref{fig:feature_force}(b), there is no deviation in the force information 
obtained from the EEF and the worktable-type force sensor, 
and the contact force is accurately obtained from the force sensor of the EEF.
On the other hand, Data2 and Data3 include deviations between the force information obtained from the two sensors, 
and they are datasets to verify the robustness against deviations (e.g., hysteresis, drift, and offset) in the force information.
In this study, we reproduced the deviation in the force information by offset. In Data2, the offset is programmatically shifted by 2~N. 
In Data3, the offset is shifted by applying a 2~N weight next to the offset acquisition.
Data4 is a dataset of linear movement in the X-Y plane after contact, 
where the origin is the contact point and the translation is 2~cm in the X-axis direction and $-$3~cm in the Y-axis direction.
This dataset is used to verify how the vibration due to the lateral movement and the dynamic friction between the tool and the grinded surface affect the accuracy. 
Table~\ref{tb:TEST} presents the combination of the training and test data.
In each case, the training data are set to Data1.
In other words, the conditions (b), (c), and (d) in Table~\ref{tb:TEST} test how well the model trained 
on Data1 with simple grinding data can maintain its performance under error-prone conditions. 
}

\textcolor{black}{
Because the robot trajectory generation is highly reproducible and 
the number of data points that can be obtained is not large, the 
collection of datasets under the same conditions will only yield 
the relationship between the specific frequency components and the 
force information. 
Therefore, we input five patterns of commands 
in the Z-axis direction of $\bm{p}^{\rm{cmd}}$: 0, $\pm1$, and $\pm2$~mm.
For each command, we obtained 15 and 5 data points for training and test, respectively.
Therefore, each dataset of Data1-4 comprises 75 training data points and 25 test data points. 
Since we focused on the 1D contact force in this study, only the contact force in the Z-axis direction was obtained.
}

\begin{table}[t]
\centering
\caption{Type of Dataset}
\begin{tabular}{c|l|l}
\hline
  &Offset error&Motion in the X-Y plane\\ \hline \hline
Data1&Not given &None\\ \hline
Data2&Given by programing&None\\ \hline
Data3&Given by weight&None\\ \hline
Data4&Not given &$x=2$cm, $y=-3$cm\\ \hline
\end{tabular}
\label{tb:Dataset}
\end{table}

\begin{table}[t]
\centering
\caption{\textcolor{black}{Type of Dataset for Training and Test}}
\begin{tabular}{c|c|c|c}
\hline
  &Training&Test&Details to verify effect\\ \hline \hline
a)&Data1&Data1&-\\ \hline
b)&Data1&Data2&Drift of the force sensor\\ \hline
c)&Data1&Data3&Drift of the force sensor\\ \hline
d)&Data1&Data4&Motion of X-Y plane\\ \hline
\end{tabular}
\label{tb:TEST}
\end{table}

\subsection{Performance Evaluation}
In this study, we verified the effectiveness of using MS as 
a feature in the estimation of the oscillatory force information during grinding. 
For this purpose, we also compared the results with those obtained using 
the force information. 

\textcolor{black}{
To reduce the amount of data 
to be inputted while matching the time-series length, 
the sampling rate was varied from 1 ms to 2 ms, 
and the input data were reduced to 256 samples (512 ms).
}

\textcolor{black}{
To compare the prediction accuracy of the model with previous models, 
we used first-order LPF and feed-forward NN (FNN) as baseline methods in addition to the CNN.
The LPF is verified as a widely used method.
The FNN has also been used in previous studies \cite{Kim2020} for accuracy improvement.
In this study, the model is constructed to be optimal for the task
and comprises two hidden layers with 50 nodes and an output layer with one node.
In the FNN+MS model, the 45$\times$17 MS time series data were converted to 755$\times$1 1D information as input.
ReLU was used as the activation function.
In each model, the mean squared error (MSE) was used as the loss function.
The learning time was set to 1000 epochs for each model.
}

Table \ref{tb:INFLUENCE} compares the root MSEs (RMSEs) of the inference 
results for each model. From a), the NN is found to be more effective in 
removing periodic vibration noise with a higher performance than the LPF, 
and the CNN is suitable for the model. Comparing the results of NN+MS with 
those of the NN alone in a), it can be confirmed that the estimation accuracy of the NN 
alone is higher. This result indicates that the performance slightly degraded 
because the low-frequency component that directly corresponded to the 
absolute value of the external force was not used. 
Note that the low-frequency information was removed in the MS, which used only the 
information from 40 Hz to 400 Hz. The results in b) and c) show 
that the FNN and CNN were slightly superior to the LPF and that adding MS 
as preprocessing significantly improved the performance. These results 
suggest that a frequency analysis of the force information is effective 
in the event of drift, because the use of low-frequency response degrades the 
accuracy in environments that are prone to drift. 
From the results in d), it can be confirmed that the performance of the 
proposed method degraded during horizontal movement. 
Because the state of contact changes during the movement, the 
relationship between the frequency and the force fluctuates, 
which affects the force estimation.

\textcolor{black}{
The effect on the estimated value is determined 
by the drift value of the sensor and the noise due to movement.
Although the drift value can be designed, the noise generated 
during the movement is determined by the material of the grinded surface, 
the grinding tool, and the movement speed of the robot.
Therefore, it is difficult to determine the effect on the estimated value in advance.
The proposed method is applicable when the position command and other parameters are adjusted 
so that the drift of the sensor is greater than the effect of noise due to lateral motion.
}

\textcolor{black}{
Table \ref{tb:low_freq_del} shows the RMSE [N] when the low-frequency dimension is removed.
Here, “None” represents that the low-frequency dimension is not deleted, 
and “1-5 dim.” represents that the lower 1-5 dimensions are deleted.
Table \ref{tb:low_freq_del} shows that the RMSEs in c) for “1-5dim.” are improved compared to “None”; 
thus, the removal of the low-frequency dimension contributes to the reduction of the error due to hysteresis.
The dimension of the hysteresis that is affected depends on the task. 
Therefore, the number of dimensions of the low frequency to be removed 
should be designed depending on the task.
}

\begin{figure}[t]
  \centering
    \includegraphics[width=7.4cm]{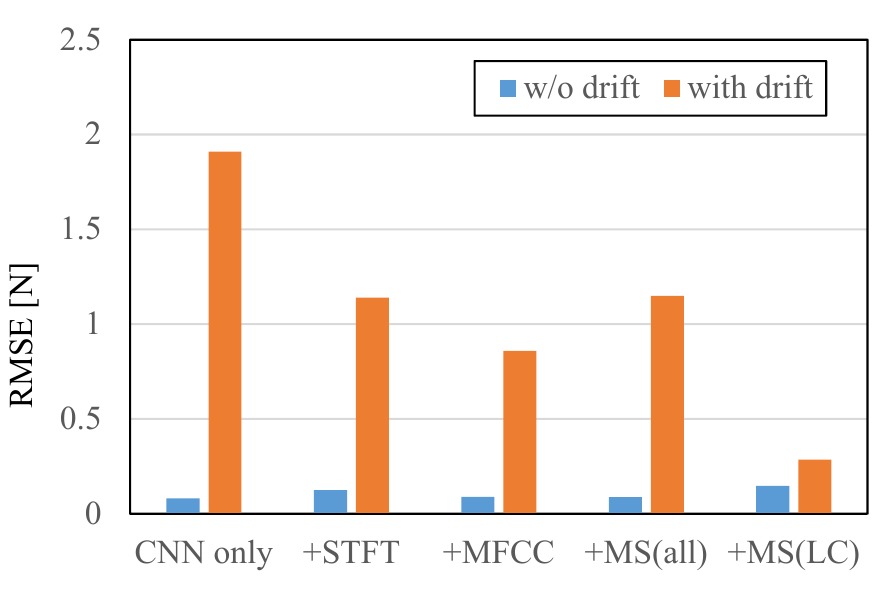}
    \caption{Comparison of frequency analysis methods.}
    \label{fig:cnncomp}
\end{figure}

To compare the frequency analysis methods, an evaluation 
test for the force estimation performance was conducted. Fig.~\ref{fig:cnncomp} shows the results are. 
Here, all the results except of the CNN only 
are for the combination with CNN. MS(all) represents the case where 
all the information up to the 50th order is used, and MS(LC) represents 
the case where the components up to the 5th order are removed.

A high force estimation accuracy was obtained using only the CNN estimation, 
and when the frequency analysis was added, the performance was 
largely the same in all cases. However, when a drift of 2~N occurred, the drift 
significantly increased the force estimation error. When the STFT, which is one of the simplest 
frequency analysis methods, was added to the CNN, its error was reduced, 
and similar results were obtained for the MFCC and MS.
These methods reduced the error because the frequency analysis method 
quantified the change in the frequency depending on the magnitude of the load 
and sent it to the CNN. However, even when the frequency analysis is implemented, 
the low-frequency component is affected by drift, and regression methods based 
on the values affected by drift will generate errors. The fact that the MS(LC) results 
outperformed the other methods in tests with drift is an indication of the effect 
of removing components up to the 5th order, which are strongly affected by drift. 
In addition, this result shows the ability to estimate the external force even with 
only the high-frequency component, which removes the absolute value of the force. 
Since the low-frequency component, which is directly related to the external force, 
is removed in MS(LC), the error is slightly increased compared with those of the other 
methods when there is no drift; nevertheless, it is small compared to the effect of drift.

\begin{table}[t]
\centering
\caption{\textcolor{black}{Comparison of RMSE [N]}}
\begin{tabular}{c|c|c|c|c|c}
\hline
  &LPF(5Hz)&FNN&CNN&FNN+MS&CNN+MS \\ \hline \hline
a)&0.153&0.107&0.081&0.141&0.147\\ \hline
b)&1.93&1.49&1.48&0.262&0.272\\ \hline
c)&1.91&1.57&1.56&0.285&0.286\\ \hline
d)&0.478&0.376&0.369&0.712&0.655\\ \hline
\end{tabular}
\label{tb:INFLUENCE}
\end{table}

\begin{table}[t]
\centering
\caption{\textcolor{black}{RMSE [N] when Low-Frequency Features are Removed.}}
\begin{tabular}{c|c|c|c|c|c|c}
\hline
  &None&1 dim.&2 dim.&3 dim.&4 dim.&5 dim. \\ \hline \hline
a)&0.074&0.143&0.170&0.165&0.151&0.147\\ \hline
c)&1.94&0.292&0.275&0.287&0.289&0.286\\ \hline
\end{tabular}
\label{tb:low_freq_del}
\end{table}

\subsection{Verification of grinding performance}
\begin{figure}[t]
  \centering
    \includegraphics[width=8.3cm]{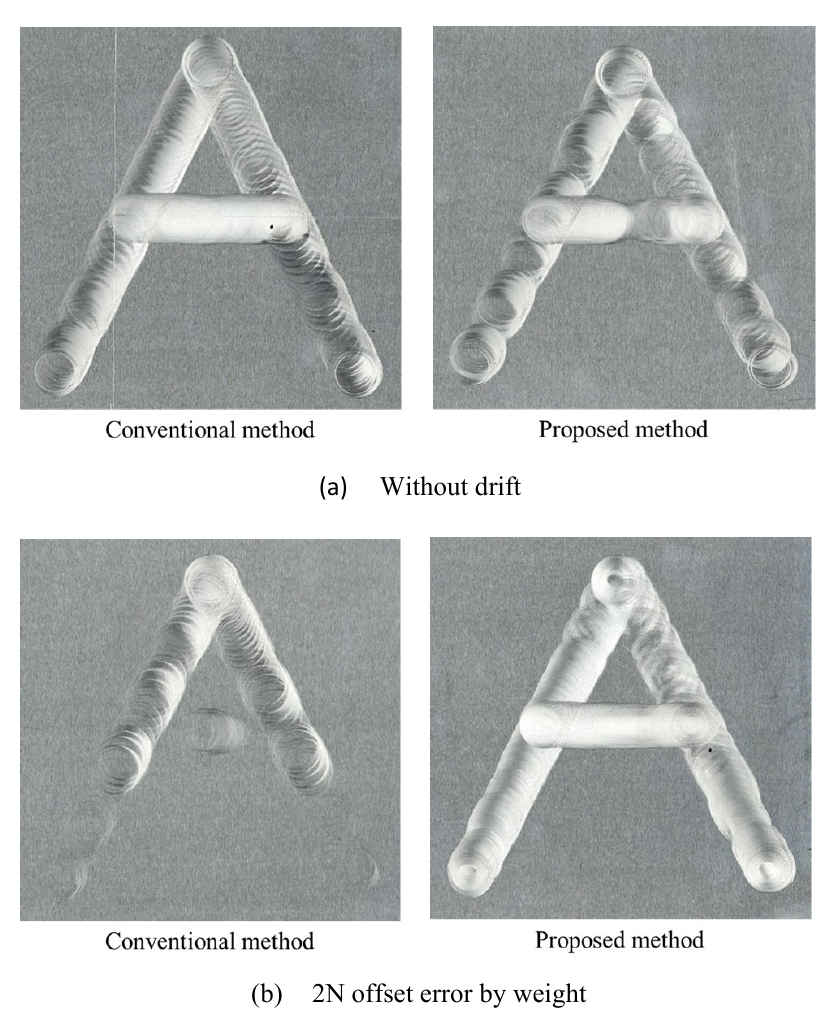}
    \caption{Results of grinding “A” task.}
    \label{fig:drawing_a}
\end{figure}

To verify the effectiveness of our proposal, we gave the robot a command 
to write the letter “A” by grinding. 
The proposed method is applied as a type of filter to the output value of the 
force sensor. 
We also conducted experiments under the condition that there is 
a deviation in the initial offset to verify the robustness against drift. 
The results, shown in Fig.~\ref{fig:drawing_a}, indicate that the quality of the 
grinding work degrades when an offset error occurs in the conventional 
method, whereas it is only slightly degraded in the case of the proposed 
method. This result confirms that the proposed method enables robust force control against drift.

\section{conclusions}\label{conclusion}
In operations with constant vibration, such as grinding and cutting, 
the vibration frequency has a strong correlation with the external force. Focusing 
on this characteristic, we proposed an external force estimation method 
based on the frequency analysis of the force information obtained from a force sensor. 
The frequency information is found to be more effective for accurate force estimation 
than the amplitude in operation with large noise caused by vibration. Moreover, it 
is experimentally demonstrated that the proposed method is more robust 
against force sensor-specific problems such as hysteresis.

The proposed method is limited to tasks that have a 
source of vibration at the hand or the contact object, such as grinding or 
cutting. Since the natural vibration depends on the load of the 
contact, and its value is always influenced by the force at the hand, in 
principle, it can be implemented without a large source of vibration. Future 
work will be to extend this method to more robots by realizing an algorithm 
that can accurately estimate the external force from minute vibrations.

\end{document}